\documentclass[a4paper, 10pt, 5p, preprint, twocolumn]{elsarticle}
\usepackage{amsmath,graphicx,color}
\usepackage[normalem]{ulem}
\usepackage[T1]{fontenc}
\usepackage{amssymb}
\usepackage{xcolor}
\usepackage{hyperref}
\usepackage{mathtools, cuted}
\usepackage{algpseudocode}
\usepackage[ruled,vlined]{algorithm2e}

\usepackage{mwe}
\usepackage{placeins}

\journal{PLB 
}

\newcommand{\xt}{\mathbf{x}}

\newcommand{\trento}{T$\mathrel{\protect\raisebox{-2.1pt}{R}}$ENTo}

\begin{document}

\begin{frontmatter}

\title{Intermediate mass dileptons as pre-equilibrium probes in heavy ion collisions}

\author[irfu]{Maurice Coquet}
\cortext[mycorrespondingauthor]{Corresponding author}
\ead{maurice.louis.coquet@cern.ch}
\author[bielefeld]{Xiaojian Du}
\author[ipht]{Jean-Yves Ollitrault}
\author[bielefeld]{S\"oren Schlichting}
\author[irfu]{Michael Winn}

\address[irfu]{Universit\'e Paris-Saclay, Centre d’Etudes de Saclay (CEA), IRFU, IRFU, D\'epartement de Physique Nucl\'eaire (DPhN), Saclay, France}
\address[bielefeld]{Fakult\"at f\"ur Physik, Universit\"at Bielefeld, D-33615 Bielefeld, Germany}
\address[ipht]{Universit\'e Paris Saclay, CNRS, CEA, Institut de physique th\'eorique, 91191 Gif-sur-Yvette, France}

\begin{abstract}
The production of dileptons with an invariant mass in the range $1~\text{GeV}<M<5$~GeV provides unique insight into the approach to thermal equilibrium in ultrarelativistic nucleus-nucleus collisions. 
In this mass range, they are produced through the annihilation of quark-antiquark pairs in the early stages of the collision. 
They are sensitive to the anisotropy of the quark momentum distribution, and also to the quark abundance, which is expected to be underpopulated relative to thermal equilibrium. 
We take into account both effects based on recent theoretical developments in QCD kinetic theory. 
We argue that the dilepton mass spectrum provides a measure of the shear viscosity to entropy ratio that controls the equilibration time.
We evaluate the background from the Drell-Yan process and argue that future detector developments can suppress the additional background from semileptonic decays of heavy flavors.
\end{abstract}


\end{frontmatter}


\section{Introduction}\label{intro}

Ultrarelativistic nucleus-nucleus collisions at the Large Hadron Collider (LHC) produce a rapidly-expanding plasma of quarks and gluons. 
By now there is ample experimental evidence that this plasma reaches a state sufficiently close to thermal equilibrium to be described by relativistic viscous hydrodynamics~\cite{Nagle:2018nvi,Busza:2018rrf,Braun-Munzinger:2015hba} and recent developments in QCD kinetic theory provide a solid theoretical basis for understanding how thermalization is achieved~\cite{Schlichting:2019abc,Berges:2020fwq}. 
However, experimental signatures of the  thermalization process itself have remained elusive so far. 
Evidence for thermal equilibration largely relies on analyses involving hadrons, which are emitted at the end of the expansion. They reflect the thermodynamic properties of the quark-gluon plasma at a temperature $T\sim 220$~MeV~\cite{Gardim:2019xjs}, long after thermalization has been achieved. 
Global Bayesian analyses~\cite{JETSCAPE:2020mzn} confirm that the constraining power of hadronic observables degrades at temperatures above 250~MeV, even though significantly higher temperatures should be reached during the early stages of the collision. 
We show that dileptons in the invariant mass range $1~\text{GeV}<M<5$~GeV, which are produced early on, provide a window to study these higher temperatures and the onset of thermalization~\cite{Martinez:2008di}, in particular the equilibration of quark abundances. 
We argue that their measurement is within reach in the next decade. 

Unlike hadronic observables, electromagnetic observables (photons and dileptons) carry information about the different stages of the evolution. Photons and dileptons are created by fluctuations of electromagnetic currents in the plasma throughout the collision process. Since they do not interact through the strong interaction, they traverse the plasma and typically reach the detector unscathed.
In comparison to photons, dileptons (either $e^+e^-$ or $\mu^+\mu^-$ pairs) are more versatile probes because they carry an additional degree of freedom, the invariant mass $M$ of the pair. 
Different mass ranges typically correspond to different production times and thus probe different stages of the system~\cite{Rapp:2013nxa}.

We focus on dilepton production with $M>1$~GeV, significantly larger than the highest temperature achieved in a central Pb+Pb collision at the LHC, which typically does not exceed 400~MeV~\cite{Hanus:2019fnc}.
In this regime where $M\gg T$, the thermal production rate is exponentially suppressed by a Boltzmann factor $\exp(-M/T)$, so that the dominant contributions come from the highest temperatures/energy densities reached in the collision. Beyond temperatures $\sim 155$~MeV, the thermodynamic state of QCD matter is a deconfined quark-gluon plasma, in which dileptons are produced through quark-antiquark annihilation. Since the highest temperatures/energy densities are achieved at early times, one expects a sizable contribution to dilepton production before thermalization is achieved.
It is therefore essential to model the pre-equilibrium dynamics. 

With regards to describing the production of electromagnetic probes during the pre-equilibrium stage, two important effects must be taken into account. 
First of all, the initial stage is expected to be highly gluon dominated~\cite{Lappi:2006fp}, and it takes time for quarks and antiquarks to be produced and reach thermal abundances~\cite{Kurkela:2018xxd,Kurkela:2018oqw,Du:2020zqg,Du:2020dvp}.
Secondly, due to the rapid longitudinal expansion, the momentum distribution of quarks is strongly anisotropic at early times~\cite{Martinez:2008di}, with the typical transverse momenta larger than longitudinal momenta. 
This has a direct effect on dilepton production, since the dominant kinematics in the limit $M\gg T$ is a head-on collision between a quark and an antiquark of opposite momenta, each carrying a momentum $\sim M/2$. 
Dilepton production probes the tail of the quark momentum distribution, and is highly sensitive to its anisotropy. 

One of the main interests of the hydrodynamic modelization is that it is robust and universal, in the sense that the medium properties enter through a small number of parameters (equation of state, transport coefficients)~\cite{Gale:2013da,Romatschke:2017ejr}. 
A major theoretical advance in recent years is the recognition that there is universality also in the pre-equilibrium dynamics. 
The breakthrough was the observation that different initializations of the non-equilibrium dynamics quickly converge to the same attractor solution~\cite{Heller:2015dha}. 
This attractor behaviour, which was discovered in the context of strong-coupling calculations, was then also identified in the weak-coupling limit (kinetic theory), first in the relaxation-time approximation~\cite{Heller:2016rtz,Blaizot:2017ucy,Strickland:2017kux} and finally in QCD kinetic theory~\cite{Kurkela:2018wud,Kurkela:2018vqr}.
Furthermore, these different modelizations lead to very similar attractors~\cite{Giacalone:2019ldn}, which paves the way to a robust modeling of the pre-equilibrium dynamics, where the information about the thermalization is encoded into a single parameter, typically the viscosity over entropy ratio $\eta/s$. 

We model dilepton production in the intermediate mass range by exploiting these new developments.
We use a simplified hydrodynamic modelization, in which the density is assumed to be uniformly distributed in the transverse plane. 
But we implement a state-of-the-art treatment of pre-equilibrium effects~\cite{Du:2020zqg,Du:2020dvp} (Sec.~\ref{s:expansion}), which is essential since, as written above, dileptons probe the tail of the quark distribution. 
Our results are presented in Sec.~\ref{s:results}, where we emphasize the dependence on the viscosity over entropy ratio, and the effect of the quark suppression in the pre-equilibrium stage. 
In Sec.~\ref{s:backgrounds}, we discuss other sources of dilepton production in this invariant mass range, which are backgrounds for the pre-equilibrium and thermal dileptons. Specifically, we estimate the direct production from the Drell-Yan process, and comment on the separation from weak decays of charmed hadrons. 
We do not discuss the background from charmonium decays, which is of interest in its own~\cite{Brambilla:2010cs,Andronic:2015wma} and results in well-identified peaks in the mass spectrum. 
Our results are summarized in Sec.~\ref{s:conclusion}, where we illustrate how the thermalization process can be constrained from the measured dilepton spectrum. 

\section{Dilepton production in a non-equilibrium QGP}
\label{s:production}
Dileptons are produced through quark-antiquark annihilation in the plasma. 
We denote by ${\bf p}_1$ and ${\bf p}_2$ the momenta of the incoming quark and antiquark, and by $P_1$ and $P_2$ their 4-momenta. 
We neglect quark masses, so that $P_i=( p_i,{\bf p}_i)$. 
The 4-momentum of the dilepton is $K=P_1+P_2$, and its invariant mass is $M=\sqrt{K^\mu K_\mu}$. 
The production rate is given by~\cite{Martinez:2008di,Kasmaei:2018oag}
\begin{strip}
\begin{equation}
\label{rate}
    \frac{dN^{l^{+}l^{-}}}{d^4x d^4 K}=\int \frac{d^3{\bf p}_1}{(2\pi)^3}\frac{d^3{\bf p}_2}{(2\pi)^3}4N_c\sum_f f_q(x,{\bf p}_1) f_{\bar q}(x,{\bf p}_2) v_{q\bar q} \sigma_{q\bar q}^{l^+l^-}\delta^{(4)}(K-P_1-P_2),
\end{equation}
\end{strip}
where $f_{q,\bar q}$ is the phase-space distribution of quarks and antiquarks, 
\begin{equation}
 v_{q\bar q} = \frac{p_1^\mu p_{2,\mu}}{p_1^0p_2^0}=\frac{M^2}{2p_1p_2}
\end{equation}
is the relative velocity between the quark and the antiquark, and 
\begin{equation}
\label{crosssection}
\sigma_{q\bar q}^{l^+l^-}=\frac{4\pi}{3} \frac{q_f^2\alpha^2}{M^2} 
\end{equation}
is the unpolarized annihilation cross section (we assume throughout this paper that the lepton mass is much smaller than $M$), where $q_f$ is the electric charge of the quark (flavours are assumed to be $f=u,d,s$ in the following) and $\alpha$ the fine-structure constant. 
The factor $4N_c$ in Eq.~(\ref{rate}) accounts for the summation over spin and colour. 
The condition of energy-momentum conservation, represented by the Dirac constraint in Eq.~(\ref{rate}), fixes the value of 
${\bf p}_2={\bf k}-{\bf p}_1$, and the angle between ${\bf p}_1$ and ${\bf k}$. 
It is therefore natural to represent ${\bf p}_1$ in a spherical coordinate system where the zenith direction is that of ${\bf k}$. 
We denote by $\theta_p$ the polar angle (angle between ${\bf p}_1$ and ${\bf k}$), and by $\varphi$ the azimuthal angle of ${\bf p}_1$, measured with respect to the plane containing ${\bf k}$ and the collision axis $z$. 
Integrating Eq.~(\ref{rate}) over ${\bf p}_2$ and $\theta_p$, one obtains:
\begin{equation}
\label{rate2}
\frac{dN^{l^{+}l^{-}}}{d^4x d^4 K}=\frac{N_c\alpha^2}{24\pi^5 k}\int_{\frac{k^0-k}{2}}^{\frac{k^0+k}{2}} dp_1 \int_{-\pi}^{\pi}d\varphi
\sum_f q_f^2 f_q(x,{\bf p}_1) f_{\bar q}(x,{\bf p}_2), 
\end{equation}
where $\cos\theta_{p}=(k^2-k_0^2+2k_0p_1)/(2kp_1)$ and ${\bf p}_2={\bf k}-{\bf p}_1$.

In thermal equilibrium, the phase-space distributions $f_q$ and $f_{\bar q}$ are independent of the directions of the momenta, so that the integrand is independent of $\varphi$. 
Using the Fermi-Dirac distribution $f(p)=1/(e^{p/T}+1)$ for $f_q$ and $f_{\bar q}$, where $T$ is the temperature of the plasma, the remaining integral over $p_1$ can be carried out analytically~\cite{Laine:2015iia}, and one obtains: 
\begin{equation}
\label{ratethermal}
\frac{dN^{l^{+}l^{-}}}{d^4x d^4 K}=\frac{N_c\alpha^2}{12\pi^4}\sum_f q_f^2 \frac{F(k)}{\exp(k^0/T)-1},
\end{equation}
where 
\begin{equation}
\label{correctionfactor}
F(k)\equiv\frac{2T}{k}\ln\left(\frac{\cosh\left(\frac{k^0+k}{4T}\right)}{\cosh\left(\frac{k^0-k}{4T}\right)}\right)
\end{equation}
is a correction factor that reduces to unity in the limit where $k^0\gg T$ and $k^0\gg k$. 
In this limit, the production rate (\ref{ratethermal}) is proportional to the Boltzmann factor $\exp(-k^0/T)$ associated with the energy of the dilepton pair. 

In this paper, we model the pre-equilibrium dynamics where the phase-space distributions are not isotropic, due to the rapid longitudinal cooling. 
We model the departure from thermal equilibrium by assuming that the phase-space distributions are functions of the variable $p_t^2+\xi^2 p_z^2$, where $p_t$ is the transverse momentum, $p_z$ the longitudinal momentum, and $\xi>1$ is the anisotropy parameter~\cite{Martinez:2008di}, whose value will be specified in Sec.~\ref{s:expansion}. 
Then, the integral over $\varphi$ in Eq.~(\ref{rate2}) is nontrivial, since the longitudinal component of the momentum depends on $\varphi$: 
\begin{equation}
\label{projectionp1}
p_{1,z}=p_1\left(\cos\theta_p\cos\theta_k+\sin\theta_p\sin\theta_k\cos\varphi\right), 
\end{equation}
where $\theta_k$ denotes the angle between ${\bf k}$ and the $z$-axis. 
The remaining components needed for the integration are defined by $p_{2,z}=k_z-p_{1,z}$, $p_{1,t}=\sqrt{p_1^2-p_{1,z}^2}$, $p_{2,t}=\sqrt{p_2^2-p_{2,z}^2}$. 
The integrals over $\varphi$ and $p_1$ in Eq.~(\ref{rate2}) are then evaluated numerically. 

Integrating over the momentum ${\bf k}$ of the dilepton, one obtains the energy spectrum, and then the mass spectrum through the change of variables $M^2=(k^0)^2-k^2$: 
\begin{equation}
\label{rateintegrated}
  \frac{dN^{l^{+}l^{-}}}{d^4xdM}= \int_{-\infty}^{\infty} dk_z\int_0^\infty 2\pi k_t dk_t \frac{M}{\sqrt{k^2+M^2}}
  \frac{dN^{l^{+}l^{-}}}{d^4x d^4 K}.
\end{equation}
In thermal equilibrium, isotropy implies that the production rate is independent of the direction of ${\bf k}$. 
However, when pre-equilibrium dynamics is taken into account, one must integrate separately over the longitudinal and transverse momenta $k_z$ and $k_t$. 


\section{Modeling the space-time evolution of the Quark-Gluon Plasma (QGP)}
\label{s:expansion}

Dilepton pairs are produced throughout the evolution of the plasma, and the dilepton rate in Eq.~(\ref{rateintegrated}) must be integrated over the space-time history of the system to obtain the yield. 
Since we are interested in the production of intermediate mass dileptons, which are predominantly produced at early times, we neglect the transverse expansion of the plasma. 
We only take into account the longitudinal expansion, which we assume to be boost invariant~\cite{Bjorken:1982qr}. 
Boost invariance implies that the dilepton yield per unit rapidity of the dilepton equals the yield per unit rapidity of the fluid. 
We further simplify the description by treating the system as homogeneous in the transverse plane, with an area $A=\int d^2\xt$, which will be specified below.\footnote{The error resulting from these two simplifications, neglecting the transverse expansion and transverse inhomogeneity, will be estimated at the end of Sec.~\ref{s:results}.}
The only remaining non-trivial integration is that on the proper time $\tau=\sqrt{t^2-z^2}$, and the yield per unit invariant mass and rapidity $y$ is given by:\footnote{Note that the integral over $k_z$ in Eq.~(\ref{rateintegrated}) is an integral over the rapidity of the dilepton relative to the fluid. Therefore, for a given value of the rapidity $y$ of the dilepton, a range of fluid rapidities contributes to the production.}
\begin{equation}\label{yield}
    \frac{dN^{l^{+}l^{-}}}{dM dy}= A
    \int_0^\infty \tau d\tau   ~\frac{dN^{l^{+}l^{-}}}{d^4xdM}.
\end{equation}
We now explain how the dependence on proper time is modeled. 
At late times (but still early enough that the transverse expansion can be safely neglected), the plasma is locally in thermal equilibrium.
The production of intermediate mass dileptons mostly takes place at high temperatures, where the equation of state of QCD is approximately conformal.
We therefore assume that the energy density $e$ and the entropy density $s$ are related to the temperature $T$ through:
\begin{eqnarray}
\label{conformaleos}
e(T)&=& \frac{\pi^2}{30}\nu_{\rm eff}T^4\cr
s(T)&=& \frac{4\pi^2}{90}\nu_{\rm eff}T^3,
\end{eqnarray}
where $\nu_{\rm eff}\approx 32$ denotes the effective number of (bosonic) degrees of freedom at temperatures in the range 250-300~MeV~\cite{Borsanyi:2013bia,Bazavov:2014pvz} 
The evolution of these thermodynamic quantities as a function of the proper time $\tau$ is determined by the conservation of entropy.  
The entropy per unit rapidity $dS/dy=A\tau s(T)$ is constant, hence $\tau T^3$ is a constant at late times. 
Its value can be inferred from the measurement of the charged particle multiplicity density $dN_{\rm ch}/d\eta$, using $dS/dy\simeq   (S/N_{\rm ch})~dN_{\rm ch}/d\eta$ with $S/N_{\rm ch}=6.7$ \cite{Hanus:2019fnc}. 
One thus obtains:
\begin{equation}
\label{hydroconstant}
\tau T^3= 8.24~{\rm fm}^{-2}~\left(\frac{A}{110 \rm fm^2}\right)^{-1} \left(\frac{dN_{\rm ch}/d\eta}{1900}\right).
\end{equation}
This equation, together with Eq.~(\ref{conformaleos}), defines the evolution of the energy density $e(\tau)$ at late times. 

Equation (\ref{hydroconstant}) shows that the temperature depends on the transverse area $A$, which is therefore a key ingredient in our model calculation. 
We evaluate $A$ by running a Monte Carlo generator of initial conditions which has been tuned to experimental data, the \trento{} model~\cite{Moreland:2014oya}. 
We fix the parameters of the model as appropriate for LHC energies~\cite{Giacalone:2017dud}.
The model returns for each event an entropy density profile $s(\xt)$, where $\xt$ labels a point in the transverse plane. 
We then define the effective area $A$ by 
\begin{equation}
\label{defA}
    A\equiv \frac{\left(\int_\xt s(\xt)\right)^2}{\int_\xt s(\xt)^2},
\end{equation}
where $\int_\xt$ denotes the integral over transverse coordinates. 
Note that Eq.~(\ref{defA}) gives the correct result for a uniform density $s_0$ within an area $A$, irrespective of the shape of that area. 
We eventually average $A$ over many events in a centrality class. 
We thus obtain the values $A=104$, $96$, $71$, $54$, $41$fm$^2$  for the 0-5\%, 0-10\%, 10-20\%, 20-30\%, 30-40\% centrality intervals in Pb+Pb collisions, which are used in the calculations of in Sec.~\ref{s:results}. 

At early times, the plasma is subject to a rapid longitudinal expansion and thus unable to sustain a sizeable longitudinal  pressure, so that thermal equilibrium is lost. 
Throughout this pre-equilibrium evolution, one can still define an effective temperature $T_{\rm eff}$ from the energy density~\cite{Heller:2011ju}, by inverting Eq.~(\ref{conformaleos}):
\begin{equation}
\label{defteff}
T_{\rm eff}(\tau)\equiv \left(\frac{30}{\pi^2\nu_{\rm eff}}e(\tau)\right)^{1/4}.
\end{equation}
Despite the loss of thermal equilibrium, different microscopic simulations have shown that the evolution of the energy density and of the longitudinal pressure are fairly universal when expressed as a function of the dimensionless scaling variable~\cite{Heller:2016rtz,Giacalone:2019ldn}
\begin{equation}
\label{defw}
\tilde{w}=\frac{\tau T_{\rm eff}(\tau)}{4\pi\eta/s}.
\end{equation}
Physically, $\tilde{w}$ can be understood as the ratio of the proper time $\tau$ to the thermalization time, which itself depends on $\tau$. 
Thermal equilibrium is recovered in the limit $\tilde{w}\gg 1$, while the limit $\tilde{w}\ll 1$ correspond to 
free streaming particles. 
In Eq.~(\ref{defw}), $\eta/s$ denotes the shear viscosity over entropy ratio, which is assumed constant for simplicity. 
Its magnitude determines the time it takes for the system to equilibrate. 
It is the only free parameter in our calculation.
Note that the value of $\eta/s$ has often been discussed in the context of anisotropic flow~\cite{Heinz:2013th}. 
Anisotropic flow develops at later times, and lower temperatures than those relevant for dilepton production. 
Therefore, the relevant value of $\eta/s$ for dilepton production is likely to be different, typically higher, than for anisotropic flow~\cite{Christiansen:2014ypa}.

\begin{figure}
  \begin{center}
  \rotatebox{0}{
          \includegraphics[scale=.5]{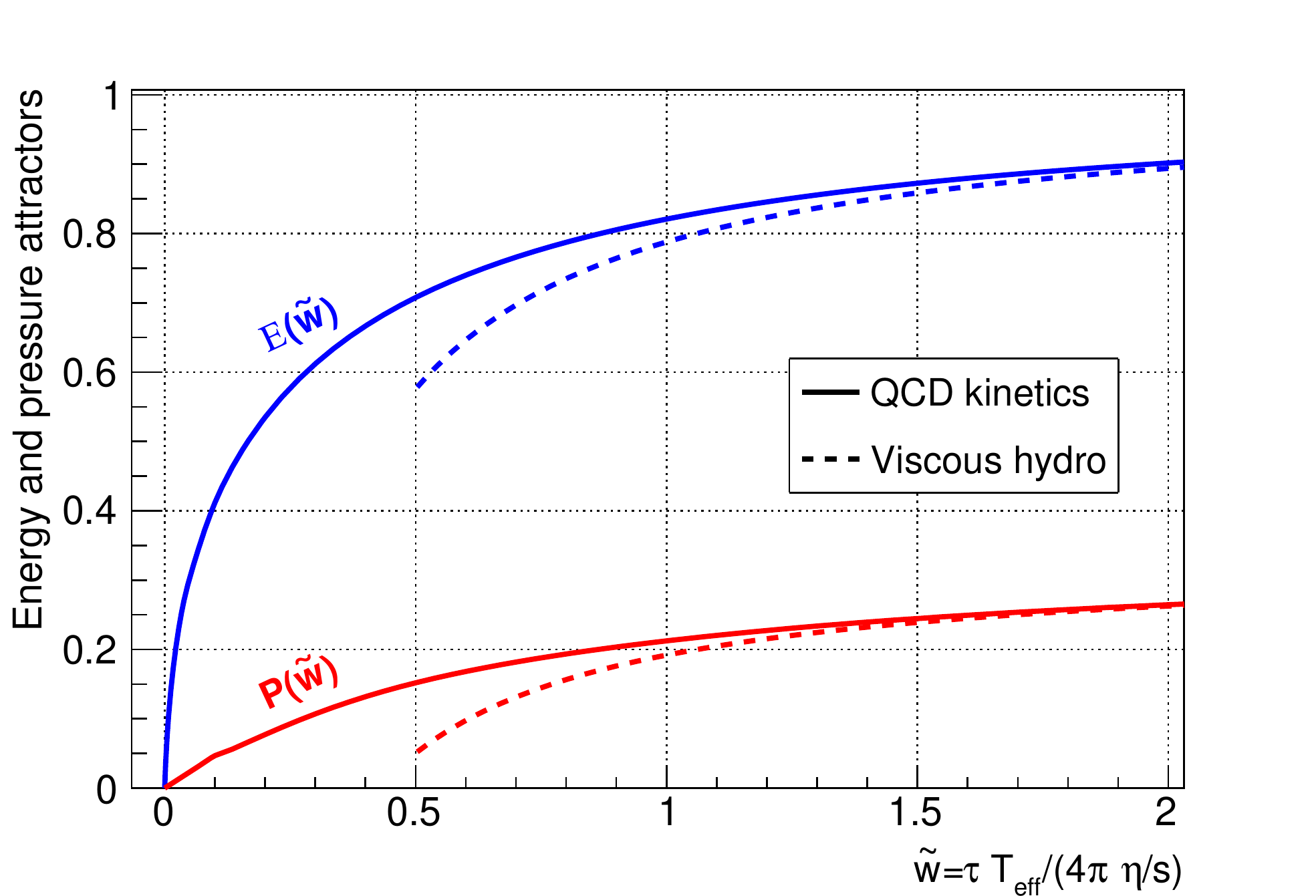}
        }
    \caption{Energy attractor $\mathcal{E}(\tilde{w})$ (blue) and the longitudinal pressure attractor $\mathcal{P}(\tilde{w})$ (red), defined by Eqs.~(\ref{energyattractor}) and (\ref{pressureattractor}), as a function of $\tilde{w}$ calculated in weak coupling regime with QCD kinetics~\cite{Du:2020dvp,Du:2020zqg}.
    Dashed lines correspond to the asymptotic behavior in (first order) viscous relativistic hydrodynamics. }
    \label{attractor}
  \end{center}
\end{figure}

One can then show~\cite{Giacalone:2019ldn} that the evolution of the energy density as a function of time is of the form
\begin{equation}
\label{energyattractor}
 e(\tau) = K\frac{\mathcal{E}(\tilde{w})}{\tau^{4/3}},
\end{equation}
where $\mathcal{E}(\tilde{w})$ is an ``energy attractor''~\cite{Giacalone:2019ldn}, which characterizes the deviation from thermal equilibrium, and $K=(\pi^2/30)\nu_{\rm eff} (\tau T^3)^{4/3}$ is a constant determined by matching the equilibrium value of Eq.~(\ref{energyattractor}) in the limit $\tilde w\to\infty$ where $\mathcal{E}(\tilde{w})\to 1$ to Eqns.~(\ref{conformaleos},\ref{hydroconstant}). 

The longitudinal pressure over energy density ratio also solely depends on the scaling variable $\tilde{w}$. 
\begin{equation}
\label{pressureattractor}
\frac{p_L(\tau)}{e(\tau)}=\mathcal{P}(\tilde{w}).
\end{equation}
The values of the energy and pressure attractors $\mathcal{E}(\tilde{w})$ and $\mathcal{P}(\tilde{w})$ 
obtained from QCD kinetic theory simulations are displayed in Fig.~\ref{attractor}.
For $\tilde{w}\gtrsim 1$, their evolution is well described by viscous hydrodynamics.
In Sec.~\ref{s:results}, we will therefore use the value of $\tilde{w}$ as a criterion to distinguish between pre-equilibrium dilepton production ($\tilde{w}<1$), and dilepton production from the hydrodynamic phase ($\tilde{w}>1$).

While the scaling functions $\mathcal{P}(\tilde{w})$ and $\mathcal{E}(\tilde{w})$ describe the macroscopic evolution of the plasma during the pre-equilibrium phase, the dilepton production rate in Eq.~(\ref{rate2}) requires information about the microscopic phase-space distribution of quarks and anti-quarks. 
While in principle these could be computed within a QCD kinetic description, it is significantly more transparent to employ an explicit parameterization which takes into account the momentum space anisotropy as well as the suppression of quarks/anti-quarks during the pre-equilibrium stage. 
Denoting the Bose-Einstein and Fermi-Dirac distribution as $f_{BE/FD}(x)=1/(e^{x}\mp1)$, we will employ the following parametrization of the out-of-equilibrium distribution
\begin{eqnarray}\label{FD}
f_{g}(\tau,p_t,p_z)
&=&f_{BE}\left(\frac{\sqrt{p_t^2 +\xi^2(\tau) p_z^2}}{\Lambda(\tau)}\right) \\
f_{q/\bar{q}}(\tau,p_t,p_z)
&=&q(\tau) f_{FD}\left(\frac{\sqrt{p_t^2 +\xi^2(\tau) p_z^2}}{\Lambda(\tau)}\right) 
\end{eqnarray}
where the anisotropy parameter $\xi(\tau) \geq 1$ characterizes the momentum space anisotropy, $0\leq q(\tau)\leq 1$  accounts for the suppression of quark/anti-quarks  and $\Lambda(\tau)$ denotes an effective transverse temperature. 

We then evaluate $\xi(\tau)$, $\Lambda(\tau)$ and $q(\tau)$ by requiring that the energy density and longitudinal pressure obtained using the ansatz in Eq.~(\ref{FD}) match those of the kinetic theory calculation.  
Explicit integration gives the following results for the contributions of quarks and gluons to the energy density and longitudinal pressure:
\begin{eqnarray}
\label{eplazhydro}
e^{(q)}(\tau) &=& q (\tau)e^{(q)}_{\rm eq}(\Lambda(\tau))~C(\xi(\tau))\;,\label{energy_q} \\
e^{(g)}(\tau) &=& e^{(g)}_{\rm eq}(\Lambda(\tau))~C(\xi(\tau))\;,\label{energy_g} \\
p_{L}^{(q)}(\tau)&=& q (\tau)e^{(q)}_{\rm eq}(\Lambda(\tau))~S(\xi(\tau))
 \\
p_{L}^{(g)}(\tau)&=&e^{(g)}_{\rm eq}(\Lambda(\tau))~S(\xi(\tau)),
\end{eqnarray}
where 
\begin{equation}
C(\xi)=\frac{1}{2}\left[\frac{1}{\xi^2} + \frac{\arctan\sqrt{\xi^2-1}}{\sqrt{\xi^2-1}}\right]
\end{equation}
\begin{equation}
S(\xi)= \frac{1}{2} \left[\frac{1}{\xi^2-\xi^4} + \frac{\arctan\sqrt{\xi^2-1}}{(\xi^2-1)^{3/2}}\right]
\end{equation}
and $e^{(q)}_{\rm eq}(T)= \frac{7\pi^2}{240} \nu_{q} T^4$ and $e^{(g)}_{\rm eq}(T)= \frac{\pi^2}{30} \nu_{g} T^4$ correspond to the equilibrium energy densities of quarks and gluons.

 \begin{figure}[ht]
  \begin{center}
    \rotatebox{0}{
         \includegraphics[scale=.48]{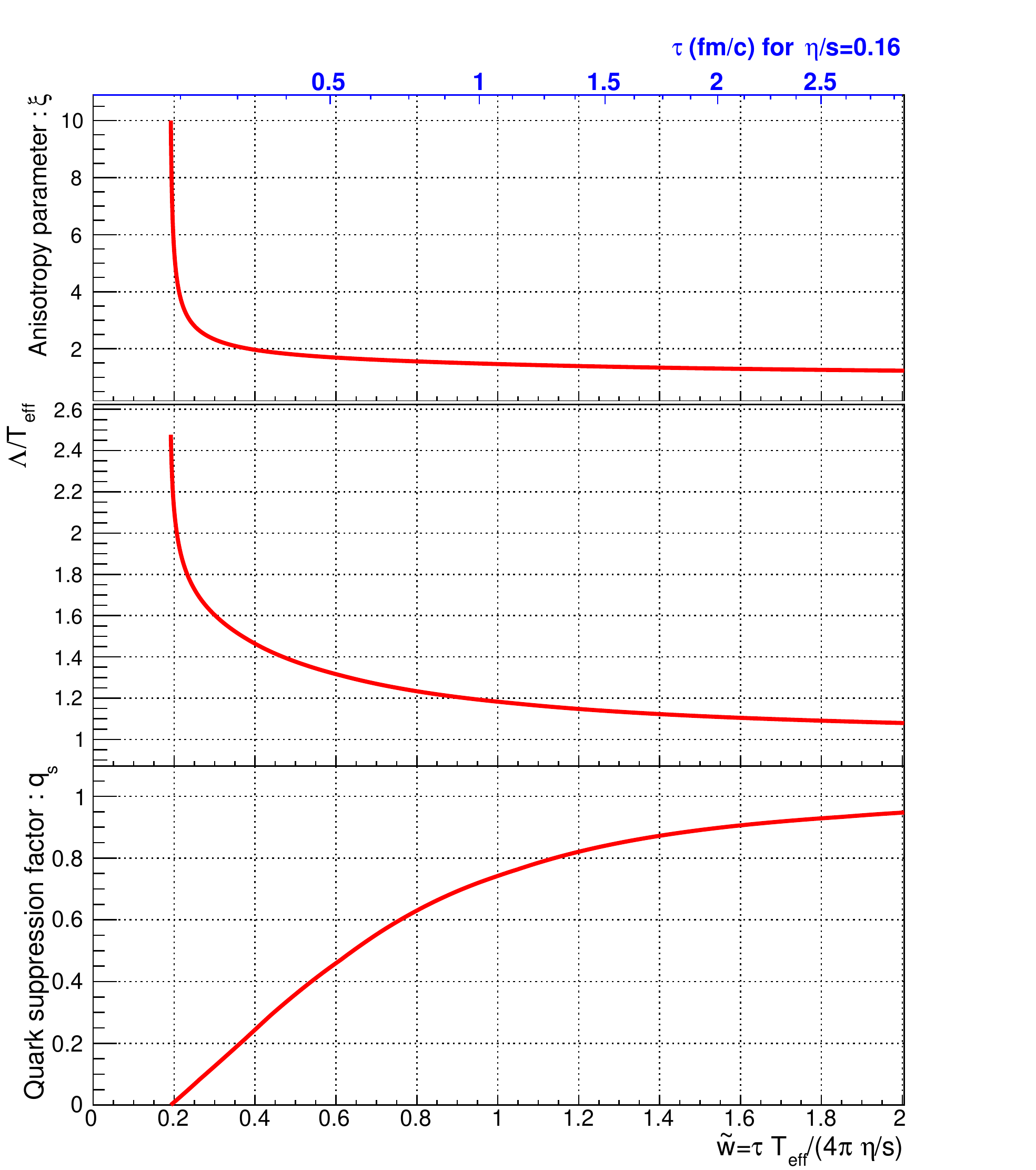}}
    \caption{(Top) Anisotropy parameters $\xi$ as a function of $\tilde{w}$, (Middle) $\Lambda/T_{\rm eff}$ as a function of $\tilde{w}$, (Bottom) quark suppression factor as a function of $\tilde{w}$.}\label{fig:all_parameters}
  \end{center}
\end{figure}

The longitudinal pressure over energy ratio (\ref{pressureattractor}) solely depends on the anisotropy parameter $\xi$: 
\begin{equation}\label{PlOverE}
\mathcal{P}(\tilde{w})=\frac{p_{L}^{(q)}(\tau)+p_{L}^{(g)}(\tau)}{e^{(q)}(\tau)+e^{(g)}(\tau)}=\frac{S(\xi)}{C(\xi)}.
\end{equation}
One obtains $\xi$ as a function of $\tilde w$ by inverting this equation. 
The quark suppression factor $q(\tau)$ is determined from the relative contributions of quarks and gluons to the overall energy density, whose dependence on $\tilde{w}$ is given by the QCD kinetic theory calculation (top panel of Fig.~31 of Ref.~\cite{Du:2020dvp}):
\begin{equation}\label{QSupp}
q(\tau)=\frac{e_{g}^{\rm eq}}{e_{q}^{\rm eq}}~\frac{e_{q}}{e_{g}}(\tilde{w}).
\end{equation}
Finally, the effective transverse temperature $\Lambda(\tau)$ is obtained by expressing the energy density $e(\tau)=e^{(q)}(\tau)+e^{(g)}(\tau)$ as a function of $\Lambda(\tau)$, using Eqs.~(\ref{eplazhydro}), and then as a function of $T_{\rm eff}(\tau)$ using 
$e(T)= (\pi^2/30)\left(\frac{7}{8}\nu_{q}+\nu_{g}\right) T_{\rm eff}^4$:\footnote{Note that this equation is not consistent with Eq.~(\ref{defteff}). 
This inconsistency is due to the fact that we need to reconcile the perturbative description of non-equilibrium effects with the non-perturbative thermodynamics of the QGP. 
We choose to evaluate the temperature $T_{\rm eff}$ using the accurate information from lattice QCD, and the ratio $\Lambda/T_{\rm eff}$ from the perturbative calculation. 
If we use the perturbative description everywhere, the temperature is $\sim 10\%$ smaller, resulting in dilepton rates smaller by a factor $\sim 2$.}
\begin{equation}
\label{TransverseTemp}
\frac{\Lambda(\tau)}{T_{\rm eff}(\tau)}= \left( \frac{\frac{7}{8}\nu_{q} +  \nu_{g}}{\left[\frac{7}{8}\nu_{q} q(\tilde{w}) + \nu_{g}\right] C(\xi(\tilde{w})}\right)^{1/4}.
\end{equation}
Fig.~\ref{fig:all_parameters} displays the variation of $\xi$, $q$, and $\Lambda/T_{\rm eff}$, defined by Eqs.~(\ref{PlOverE}), (\ref{QSupp}) and (\ref{TransverseTemp}), as a function of the scaling variable $\tilde{w}$. 
The proper time is then related to $\tilde{w}$ using Eqs.~(\ref{defteff}), (\ref{defw}) and (\ref{energyattractor}).

 \begin{figure*}
  \begin{center}
    \rotatebox{0}{
        \includegraphics[width=.49\linewidth]{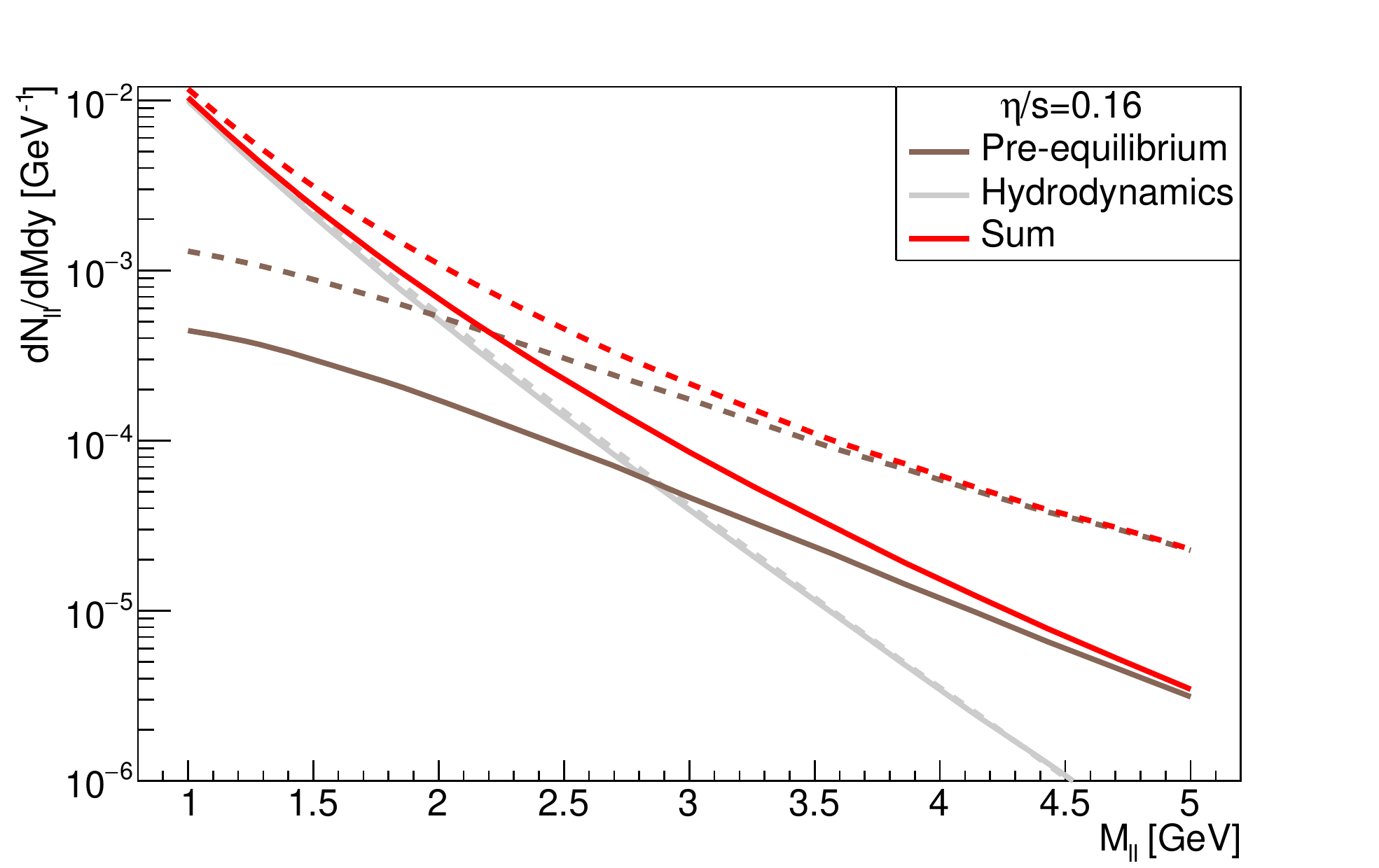} 
        \includegraphics[width=.49\linewidth]{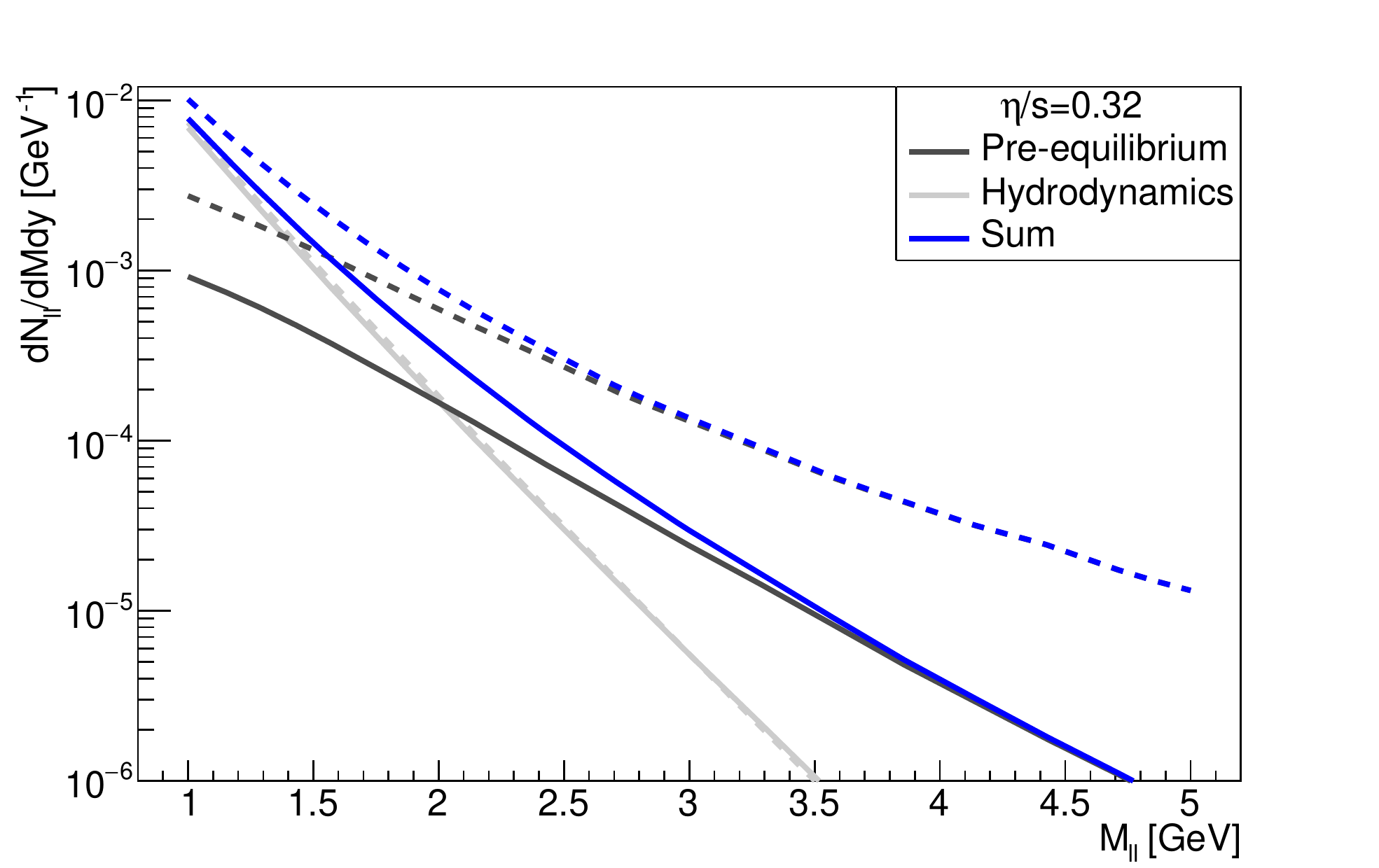}}
    \caption{Dilepton production yields $dN_{ll}/dMdY$ in the $0-5\%$ most central 5.02 TeV Pb+Pb collisions at forward rapidity $y=2$. Red (left panel) and blue (right panel) curves show the results including (full lines) and not including (dashed lines) the quark suppression factor, for shear viscosity $\eta/s=0.16,0.32$ in the left and right panels. We also show separately the contributions (see text) from the pre-equilibrium phase (dark grey) and hydrodynamic phase (light grey).}
    \label{contrib_pre_eq_supp}
  \end{center}
\end{figure*}

 \begin{figure}[ht]
  \begin{center}
    \rotatebox{0}{
        \includegraphics[width=\linewidth]{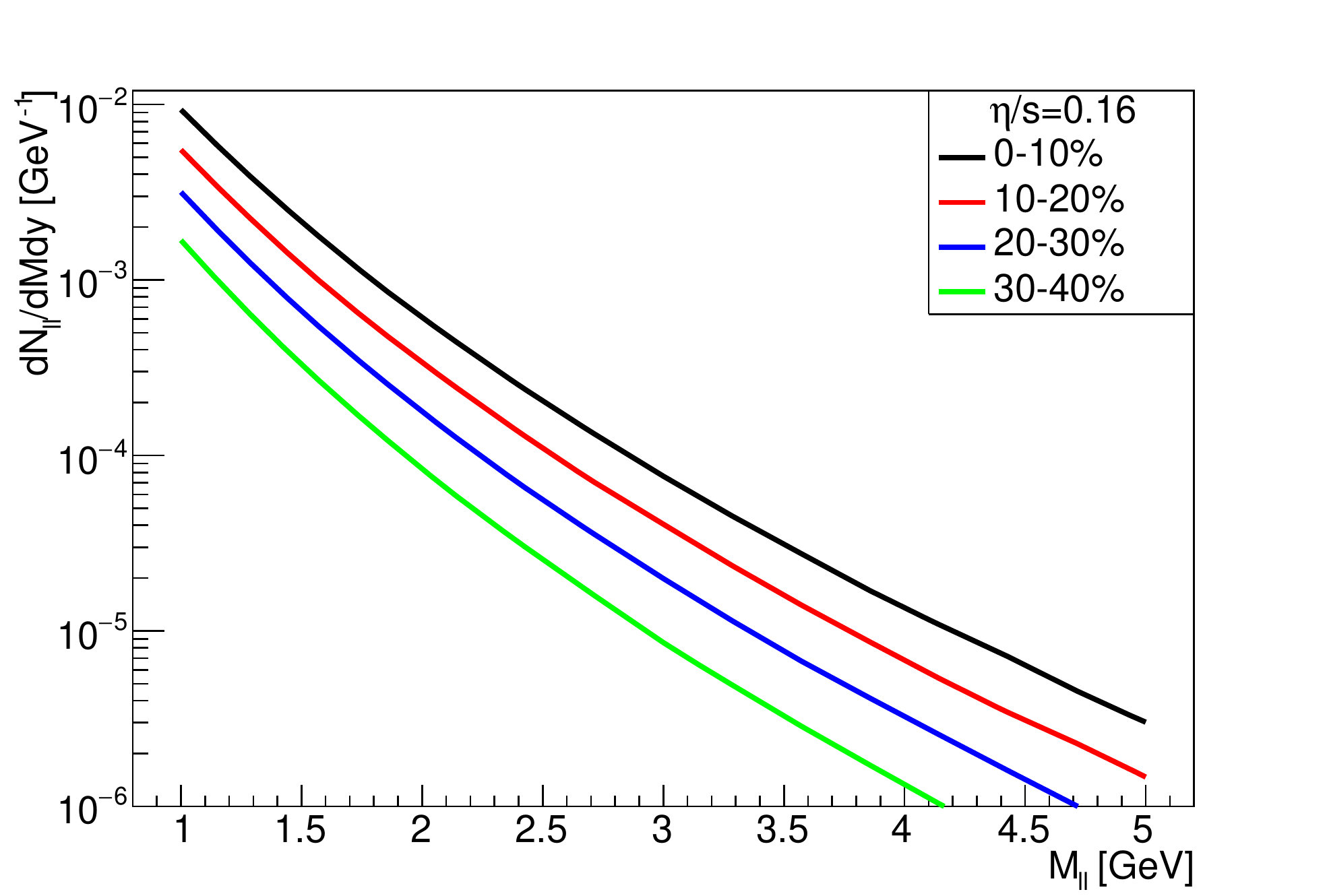} }
    \caption{Centrality dependence of the dilepton production yields $dN_{ll}/dMdY$ in 5.02 TeV Pb+Pb collisions at forward rapidity $y=2$ for shear viscosity $\eta/s=0.16$ (with quark suppression). Different centrality classes 0-10\% (black), 10-20\% (red), 20-30\% (blue) and 30-40\% (green) show are larger at suppression of dilepton production at high invariant masses. }
    \label{centrality}
  \end{center}
\end{figure}

\section{Simulation results}
\label{s:results}

In this Section, we evaluate the dilepton yield Eq.~(\ref{yield})  in Pb+Pb collisions at $\sqrt{s_{NN}}= 5.02$ TeV, and at rapidity $y=2$, corresponding to the acceptance of the LHCb experiment. 
The corresponding charged-particle multiplicity in the 0-5\% centrality bin is $dN_{ch}/d\eta\approx 1900$~\cite{Adam:2016ddh}.  
Eq.~(\ref{hydroconstant}) then gives $\tau T^3\approx 8.24$ fm$^{-2}$ in the hydrodynamic regime. 
Fig.~\ref{contrib_pre_eq_supp} displays the variation of the dilepton yield for this centrality range as a function of invariant mass, for two values of the early-time shear viscosity $\eta/s$, turning on or off quark suppression. One observes a significant dependence of the dilepton yield on both parameters, which we now explain. 


Since the system is approximately described by viscous hydrodynamics for $\tilde{w}>1$, we define $\tau_{hydro}$ by $\tilde{w}(\tau_{hydro})=1$.
Note that $\tilde{w}=1$ does not imply that the pressure is isotropic: Fig.~\ref{attractor} shows that that $P_L/e~(\tilde{w}=1)\approx 0.2$, smaller than the value $\frac{1}{3}$ corresponding to isotropy.  
The system is still slightly out-of-equilibrium at $\tau_{hydro}$, yet it is correctly modeled by viscous hydrodynamics. 
For central Pb+Pb collisions, assuming a viscosity $\eta/s=0.16$, we obtain $\tau_{hydro}\approx1$ fm/c. 
$\tau_{hydro}$ scales with viscosity like  $(\eta/s)^{3/2}$~\cite{Du:2020zqg}, and this variation explains qualitatively the dependence of the dilepton yield on $\eta/s$. 
If we lower the viscosity, the system approaches the hydrodynamic regime faster. Now, it is in this regime that the decrease of the energy density is fastest. For fixed  charged-particle multiplicity $dN_{ch}/d\eta$, lower viscosity thus implies higher initial energy density, and higher temperature throughout the out-of-equilibrium evolution. 
Higher temperature in turn implies larger dilepton yields. 
Note that a non-monotonic evolution of the dilepton yield on viscosity was observed in the three-dimensional hydrodynamic calculation of  Ref.~\cite{Ryblewski:2015hea}. This phenomenon  does not occur in our calculation, where the transverse expansion is neglected.

For the sake of illustration, we also separate in Fig.~\ref{contrib_pre_eq_supp} the contributions of pre-equilibrium and hydrodynamics to the total yield, where we define the pre-equilibrium contribution  by $\tau<\tau_{hydro}$ and the hydrodynamic contribution by $\tau>\tau_{hydro}$.
The hydrodynamic contribution dominates at lower invariant mass. 
It has little sensitivity to quark suppression, but a sizeable sensitivity to $\eta/s$.
The pre-equilibrium contribution is strongly sensitive to both $\eta/s$ and quark suppression. 
It dominates at high invariant mass. 
Note that quark suppression decreases the dilepton yield by a large factor for $M\gtrsim 3$~GeV. 
This shows that it is essential to model chemical equilibration, in addition to kinetic equilibration, in order to describe thermal dilepton production in this mass range.  

Fig.~\ref{centrality} displays the dependence of the dilepton yield on the collision centrality. 
Different centralities correspond to variations of the charged particle multiplicity $dN_{\rm ch}/d\eta$ and the transverse area $A$, resulting in different values of the constant in Eq.~(\ref{hydroconstant}). 
The variation of the dilepton yield with centrality is faster than that of the hadron multiplicity $dN_{\rm ch}/d\eta$. That is, the dilepton per charged hadron increases for central collisions~\cite{Rapp:2013nxa}. 
In the hydrodynamic regime, one expects the dilepton yield to scale typically like the space-time volume, which is proportional to $(dN_{\rm ch}/d\eta)^{4/3}$. 
This scaling explains the centrality dependence  at low invariant mass $M$, since low invariant masses originate from the hydrodynamic phase. 
For larger values of $M$, the centrality dependence is even stronger. 
The reason is that $\tau_{hydro}$ is smaller in more central collisions. 
Faster equilibration implies higher initial temperatures as explained above, and this enhances dilepton production in the pre-equilibrium phase.

We conclude this Section by estimating the uncertainties on our results from the simplifications made in Sec.~\ref{s:expansion} regarding the space-time history. 
First, we have neglected the transverse expansion, which  cools the system and results in smaller dilepton rates. 
To evaluate the effect quantitatively, we have compared our result with that of Kasmaei and Strickland~\cite{Kasmaei:2018oag} who implement a full three-dimensional viscous hydrodynamic calculation, taking into account momentum anisotropy, and neglecting  quark suppression. 
With the same setting (same initial time, $\eta/s$, centrality, $p_T$ interval, and hadron multiplicity, and neglecting quark suppression), our result for $dN/dMdy$ has the same dependence on $M$, but is a factor $\sim 2$ higher. 
Part of this discrepancy is due to the fact that they only consider the production by $u$ and $d$ quarks, while we also consider the production by $s$ quarks. 
We attribute the remaining difference, which is roughly a factor $1.7$, to the effect of the transverse expansion, which decreases the dilepton rate. 

Second, we have assumed that the transverse density profile is homogeneous. 
This is a major simplification, as we are not taking into account the fact that the density is larger in the centre of the fireball, and we are also neglecting event-to-event fluctuations, which create hot spots in the initial profile~\cite{Gyulassy:1996br}. 
High-mass dileptons typically come from these hot spots. 
We therefore expect that our calculation underestimates the dilepton yield for large $M$.
In order to evaluate this effect quantitatively, we have carried out a calculation in which we replace the uniform profile with a fluctuating profile given by a Monte Carlo Glauber calculation. 
Our results, which will be shown in a forthcoming publication, show that the simplified calculation gives the same result as the more realistic one within $\sim 10\%$ for $M\sim 1$~GeV, and underestimates the yield by a factor $1.5-2$ for $M=5$~GeV. 
The increase of the dilepton yield for large $M$, due to fluctuations, is larger for more peripheral collisions. 
It is likely that the centrality dependence will be similar for all values of $M$ once fluctuations are taken into account (and not stronger for large $M$, as found above with a uniform density). 

The important point is that quark suppression, which we include, has a much larger effect on the dilepton yield than the effects we neglect.
The uncertainty coming from the modeling of this quark suppression is likely to be larger than the errors resulting from the simplifications made in Sec.~\ref{s:expansion}. 
Modeling chemical equilibration is crucial for  dilepton production at intermediate masses. 

Finally, note that Churchill {\it et al.\/}~\cite{Churchill:2020uvk} find dilepton rates orders of magnitude smaller. 
The difference is likely due to the fact that in their calculation, the initial momentum of gluons $f(p)$ goes to 0 at high $p$. 
Now, dileptons in the considered mass range are produced by the annihilation of a high-momentum quark and a high-momentum antiquark, which are themselves produced by high-momentum gluons. 
It is therefore essential to model as realistically as possible the tail of the initial gluon  distribution. 
In addition, these authors only consider production at very early times ($\tau<0.4$~fm/c). 

\section{Backgrounds and their suppression}
\label{s:backgrounds}

 \begin{figure}[ht]
  \begin{center}
    \rotatebox{0}{
        \includegraphics[width=\linewidth]{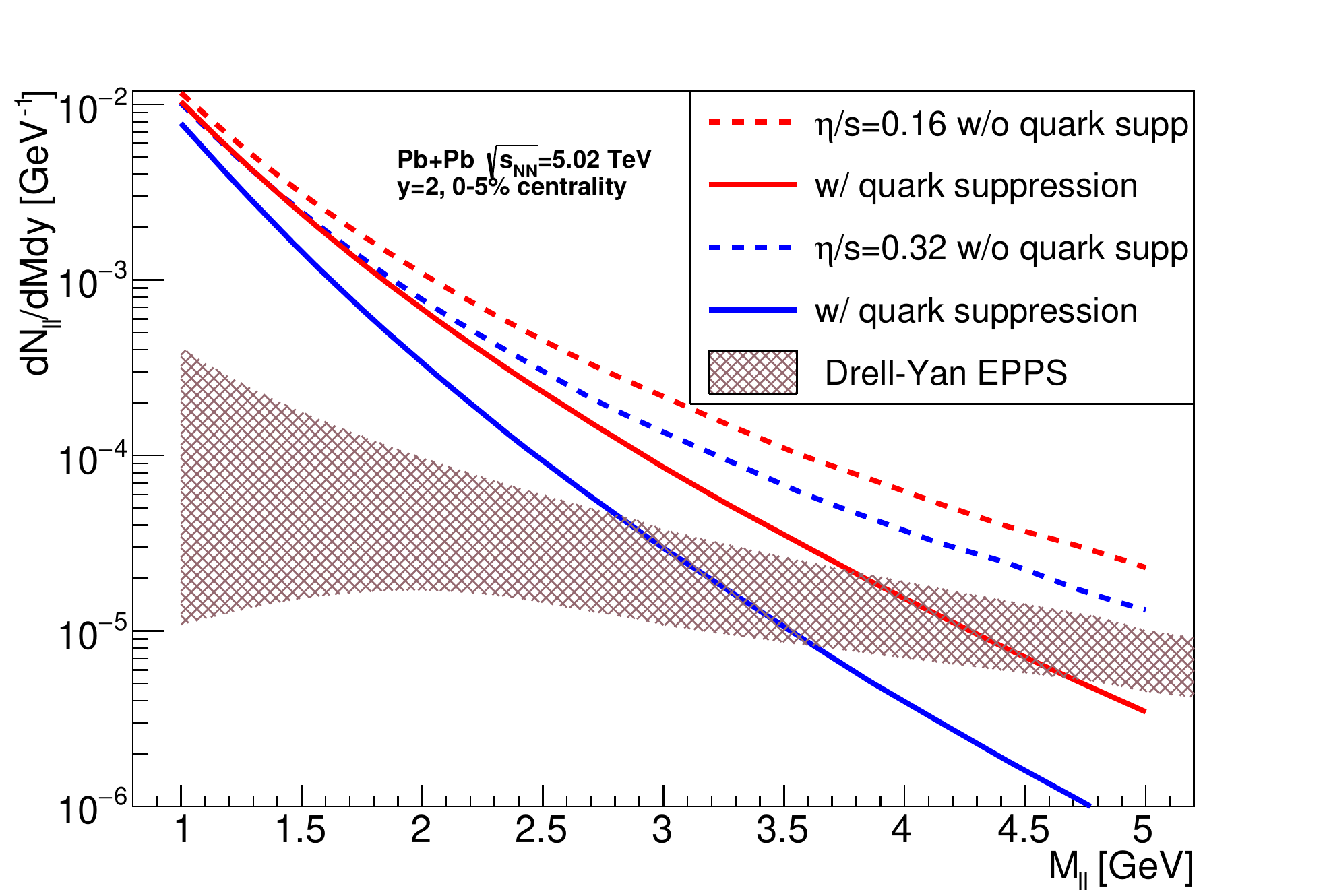}}
    \caption{Dilepton production yields in the $0-5\%$ most central 5.02 TeV Pb+Pb collisions at forward rapidity $y=2$ for different values of $\eta/s$, with and without quark suppression, from Fig.~\ref{contrib_pre_eq_supp}, compared with the Drell-Yan rate calculated at NLO with EPPS nuclear PDFs.}
    \label{fig:DY}
  \end{center}
\end{figure}

\begin{figure}[ht]
  \begin{center}
    \rotatebox{0}{
        \includegraphics[width=\linewidth]{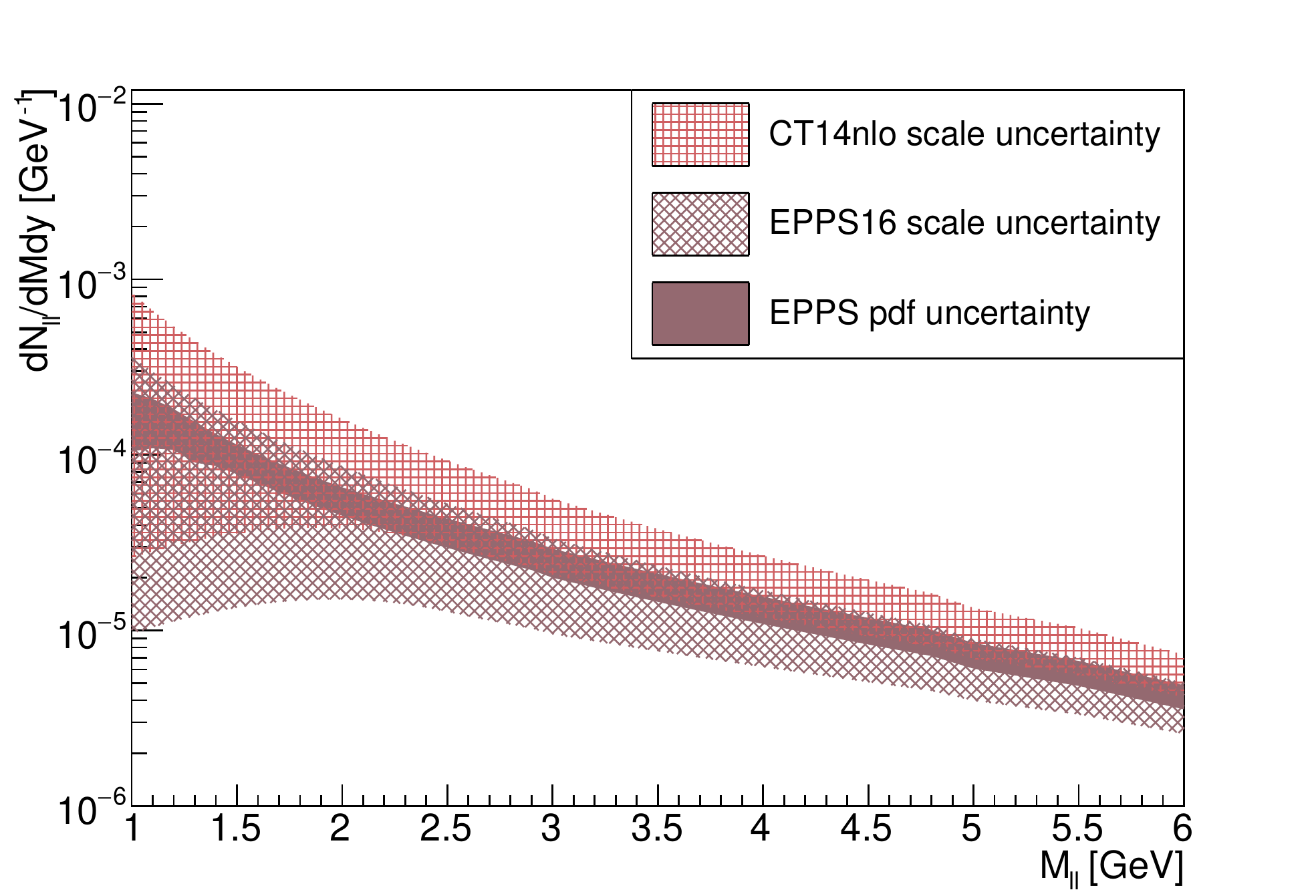}}
    \caption{Comparison of NLO calculations of Drell-Yan production in Pb+Pb collisions at $\sqrt{s_{NN}}$=5.02 TeV and $y=2$ for CT14 proton and EPPS nuclear PDFs. The crossed areas correspond to the scale uncertainty. The solid band is the uncertainty from the nPDF in the EPPS calculation. }
    \label{pp_epps}
  \end{center}
  \end{figure}

Now that we have established the signal, we comment on the known backgrounds in experimental measurements.
At large invariant masses, dilepton production in hadron-hadron collisions at the LHC is dominated by the Drell-Yan process. This background is experimentally irreducible. 
Therefore, it defines an upper bound on the mass below which thermal and pre-equilibrium dilepton production can be isolated.
Drell-Yan  production can be calculated in perturbative QCD in collinear factorization up to next-to-next-to-leading order (NNLO). 
Here we use NLO calculations, whose precision is sufficient for our purpose. 
Within collinear factorization, the uncertainties are very small at the Z-pole. They grow in the intermediate mass range due to large scale uncertainties~\cite{Eskola:2001rz}. In addition,  the nuclear parton distribution functions (nPDF) in the probed phase space are only scarcely constrained~\cite{Kovarik:2015cma,Eskola:2016oht,Walt:2019slu,AbdulKhalek:2020yuc}. 
We perform a calculation based on the EPPS nPDFs~\cite{Eskola:2016oht} and the Drell-Yan Turbo software~\cite{Camarda:2019zyx} neglecting the centrality dependence of nPDFs and assuming $T_{AA}$ scaling of the cross section in the 0-5$\%$ centrality window~\cite{Miller:2007ri}. 
The  Drell-Yan calculation is shown in Fig.~\ref{fig:DY} together with the thermal production. The shaded band corresponds to the independent variation of the factorization and renormalization scale by a factor two. 
For its upper limit, corresponding to factorization and renormalization scales equal to twice the Drell-Yan dilepton pair mass, we observe that the thermal production dominates the production below a mass $2.7$~GeV ($3.6$~GeV) for $\eta/s =0.32$ ($\eta/s =0.16$) even if we include quark suppression.  Without quark suppression, the thermal production dominates the yield up to masses above 5~GeV.   
Considering the separation between pre-equilibrium and thermal emission carried out in Fig.~\ref{contrib_pre_eq_supp}, we conclude that the pre-equilibrium emission is the dominant source of dilepton production in the mass range  $2.7 - 3.6$~GeV for $\eta/s=0.16$  ($2 - 2.7$~GeV for $\eta/s=0.32$). 

Fig.~\ref{pp_epps} displays as a dark band the uncertainty from the nPDF itself. 
We also show on this figure the calculation using the free-nucleon PDF, in order to illustrate the importance of nuclear effects for the Drell-Yan process. 

Drell-Yan dilepton pairs exhibit a different transverse momentum distribution ($p_T$) than the pre-equilibrium or thermal dilepton pairs. For the Drell-Yan pair, $p_T$ is  either given by the intrinsic $k_T$ of the incoming partons or  by the recoiling jets at higher orders, whereas for the thermal and pre-equilibrium pair, $p_T$  is given approximately by a Boltzmann distribution 
Apart from the very different slopes as a function of invariant mass, these different scales provide additional means to discriminate the production source.  
An additional source of prompt dilepton production, via two-photon scattering, has been observed in ultrarelativistic heavy-ion collisions by STAR~\cite{Adam:2018tdm} and ATLAS~\cite{Aaboud:2018eph}. This source of dileptons has a small characteristic transverse momentum scale related to the inverse transverse impact parameter 
 and their production rate drops quickly towards most central collisions~\cite{Klein:2020jom}. 
The kinematics and the dependence as function of centrality should allow to separate this contribution from the harder thermal contribution. 

\begin{figure*}
  \begin{center}
    \rotatebox{0}{
        \includegraphics[width=.49\linewidth]{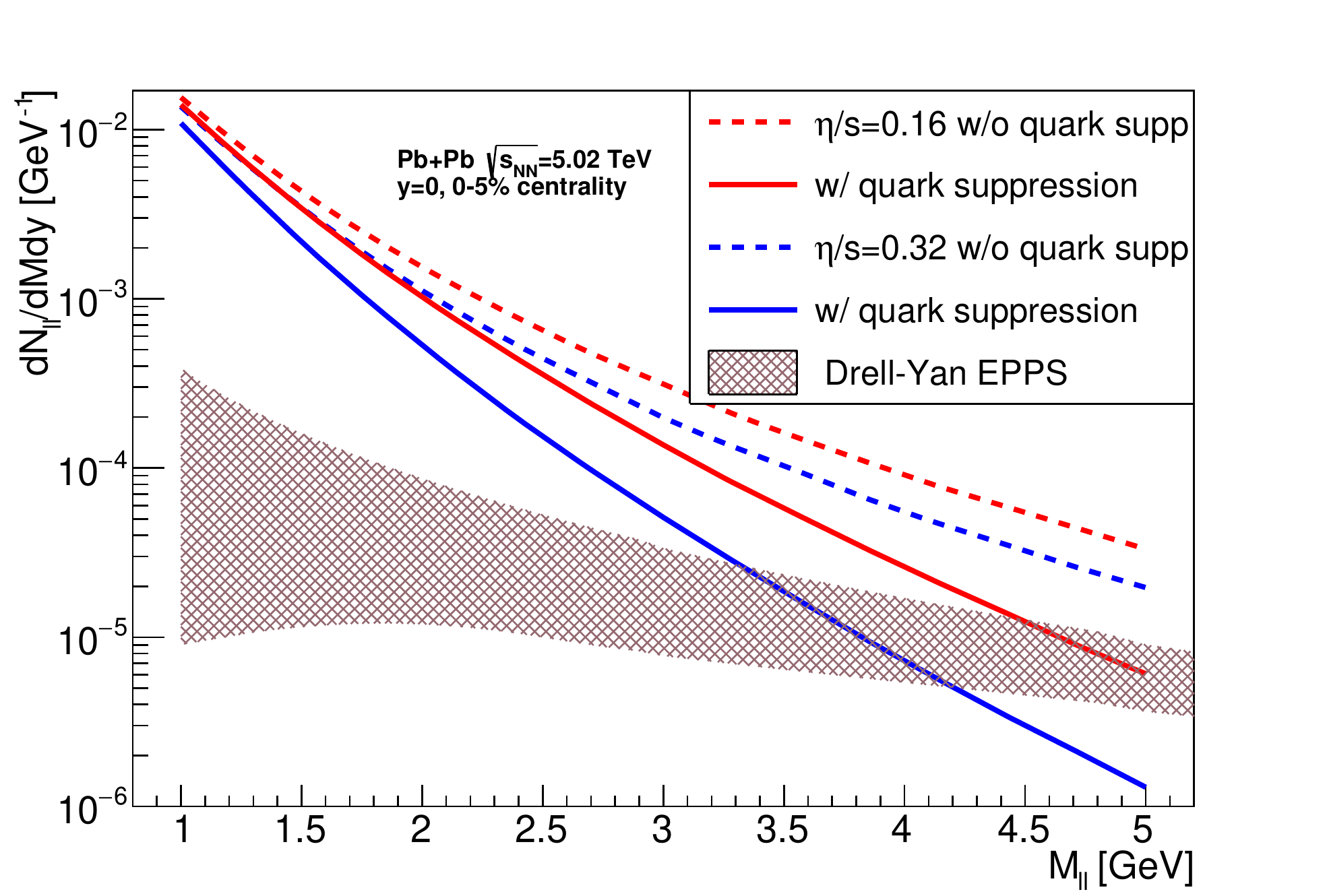}\includegraphics[width=.49\linewidth]{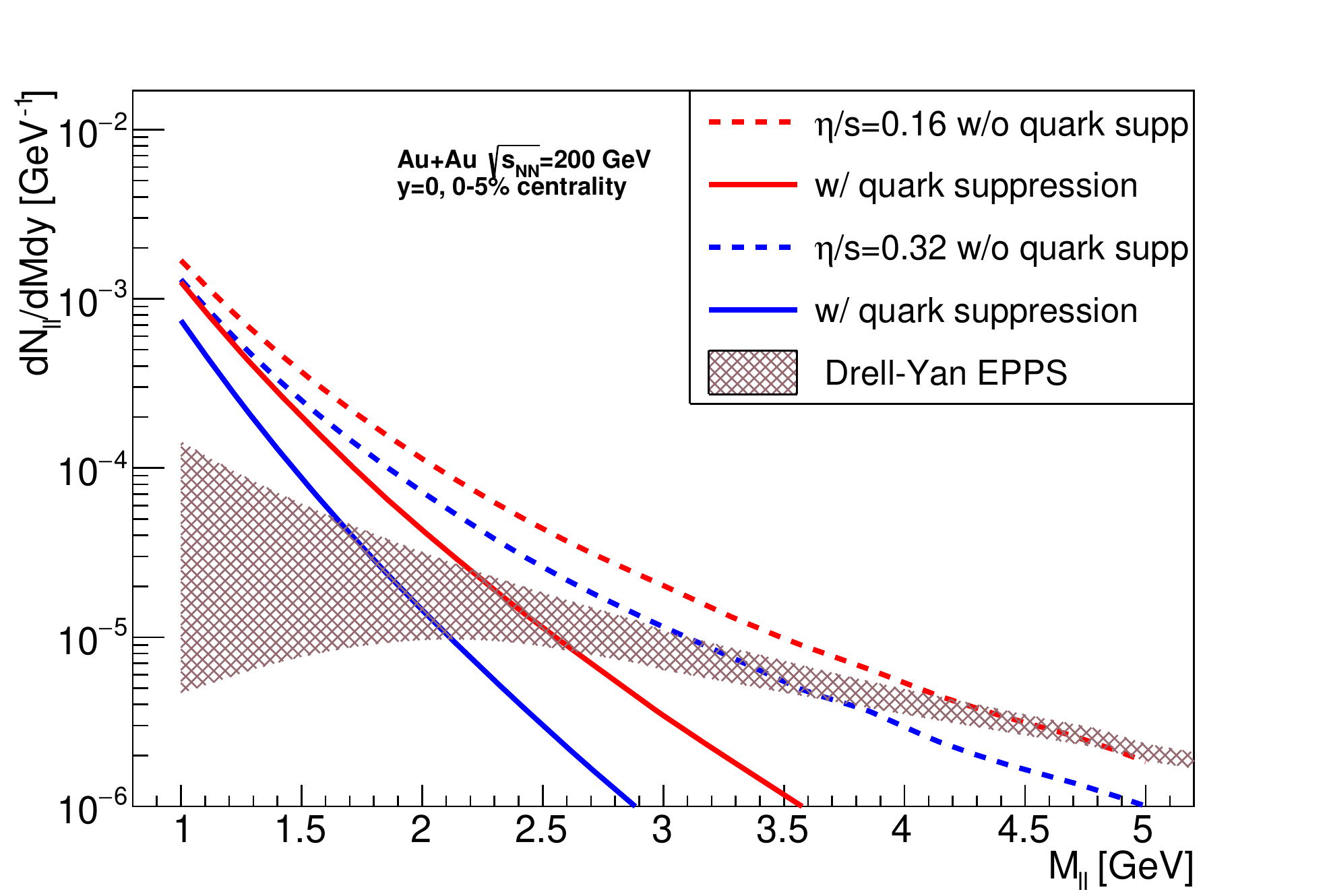}}
    \caption{Dilepton yields at $y=0$ for different values of $\eta/s$, with and without quark suppression, and Drell-Yan rate calculated at NLO with EPPS pdf, for Pb+Pb at $\sqrt{s_{NN}}=5.02$ TeV (left) and Au+Au at $\sqrt{s_{NN}}=200$ GeV (right).}
    \label{fig:ALICE}
  \end{center}
\end{figure*}
At the LHC, pairs of charm-anticharm and beauty-antibeauty quarks are produced abundantly via the strong interaction. The weak decays of the resulting charm and beauty hadrons exhibit an approximately 10$\%$ (10$\%$) probability to emit a electron/positron  (muon/antimuon). This large number of leptons yields to a large combinatorial background of lepton-antilepton pairs. This dilepton source dominates the thermal and pre-equilibrium intermediate mass dilepton production. However, the sizeable charm and beauty hadron lifetimes and their finite momentum in the laboratory lead to a sizeable displacement of the decay vertex of the hadron with respect to the primary vertex of the collision. These secondary decay vertices as well as the resulting displacement of the lepton with respect to the primary vertex are measurable with modern silicon vertex detectors very close to the vertices. The transverse distance between the lepton track and the primary vertex are already exploited in NA60~\cite{Arnaldi:2008fw} at the SPS and combinations of them in current dilepton studies for signal and background separation in ALICE~\cite{Acharya:2018ohw} at the LHC. However, studies for the ALICE upgrade for the 2030ies show that the extraction of the signal will remain systematically limited by the quantification of the background kinematics~\cite{Citron:2018lsq}. 

The LHCb experiment, so far dedicated to charm and beauty physics in proton-proton physics, is planning to exploit heavy-ion collisions in their upgrade phase 2~\cite{Bediaga:2018lhg}.  This detector has two key features: a moveable vertex detector getting as close as about 5~mm to the vertex and the forward geometry leading to a longitudinal boost of all particles in acceptance. Hence, the detector layout allows to reach low transverse momentum, key for thermal QGP and pre-equilibrium signatures, at finite momentum. First fast simulation studies indicate unprecedented intermediate mass dilepton studies with muons. LHCb has studied prompt dimuon production in Run 2 proton-proton collisions in a mass range from production threshold up to the Z boson mass, in particular in the search for dark photons with stringent momentum selections~\cite{Aaij:2017rft,Aaij:2019bvg}. In addition to the background from semi-leptonic decays of heavy flavor hadrons discussed here, the misidentification of pions as muons is an important background source in the current set-up.

Based on the same vertex detector performances as the LHCb detector of Run 3,  
we conducted rapid simulations \cite{Cowan:2016tnm} to give first estimates of the background rejection that could be achieved in a LHCb Upgrade 2 setup. Only the combinatorial background coming from the semi-leptonic decay of charmed hadrons was considered. The variables used for the rejection were the distance of closest approach (DCA) of single-track muons satisfying the selection $p_T > 0.5$ GeV with respect to the primary vertex of the interaction, and the longitudinal displacement of the secondary vertex produced by the considered semi-leptonic decay. Cutting on this last parameter we assumed that this secondary vertex was correctly identified, which is a strong assumption. Thus, we considered a conservative cut, rejecting tracks associated with a secondary vertex longitudinally displaced by more than three times the longitudinal vertex resolution expected for LHCb U2. With these assumptions and varying the cutting parameter on DCA as well as the nuclear modification factor for charm mesons $R_{AA}$ between 0.5 and 1, we estimated a signal/background from 0.3 up to 1.4 for dimuons in the mass range 1 to 3 GeV. For this first feasibility study we used our results for the dilepton yield at $y=2$ with $\eta/s=0.16$ for the $0-5\%$ most central collisions at 5.02 TeV. This estimation for the signal was compared to thermal dilepton rates computed in a fireball model provided by Ralf Rapp \cite{Rapp:1999us, vanHees:2007th, Rapp:2013nxa}, after being scaled from 2.76 TeV to 5.02 TeV based on charged-particle multiplicity. The numeric values resulting from this calculation were found to be compatible with the ones obtained in our approach.

  In addition to LHCb, the ALICE collaboration members expressed interest to build a completely new fully silicon-based detector at central rapidity with electron identification employing a vertex detector with similar performance to LHCb for the momenta in question~\cite{Adamova:2019vkf}. To this end, we plot on Fig.~\ref{fig:ALICE} our calculation for dilepton yields at midrapidity.  
  Note that in this kinematic range, the invariant mass range for which thermal production dominates over the Drell-Yan background is extended by $\sim 0.5$~GeV, in comparison with Fig.~\ref{pp_epps}. 
  Both detector systems will shed light into the chemical equilibration and the kinetic properties of the first 1 fm/c of heavy-ion collisions via dileptons. For completeness, we provide in Fig.~\ref{fig:ALICE} the calculation for central Au+Au collisions at the top collision energy of the Relativistic Heavy-Ion Collider at Brookhaven (the corresponding values of the charged particle density at mid-rapidity and of the transverse area in Eq.~(\ref{hydroconstant}) are $dN_{ch}/d\eta\simeq625$~\cite{Bearden:2001qq} and $A=100$~fm$^2$) compared with Drell-Yan production within the same set-up.  
  Note that quark suppression has an even larger effect than at the LHC. 
  Once it is taken into account, thermal dilepton production is smaller than Drell-Yan as soon as $M$ exceeds $2.2$~GeV, even if the viscosity over entropy ratio $\eta/s$ is as low as $0.16$.

\section{Conclusion and outlooks}
\label{s:conclusion}

We conclude that intermediate mass dilepton production is sensitive to the very early stages of heavy-ion collisions. The production yield is strongly sensitive to the early-stage shear viscosity over entropy density and to the chemical equilibration of the medium. 
We have carried out explicit calculations of the mass spectrum for two realistic choices of $\eta/s$. 
The results shown in Fig.~\ref{contrib_pre_eq_supp} are well fitted by the function:  
\begin{equation}
    \label{ratefit}
    \frac{dN^{l_+l_-}}{dMdy}=C\left(1+\frac{M}{nT_0}\right)^{-n},
\end{equation}
where $C\simeq 0.5$~GeV$^{-1}$, $T_0\simeq 0.2$~GeV are essentially independent of $\eta/s$, and the sole dependence on the viscosity lies in the exponent $n$:
\begin{equation}
\label{exponent}
n\simeq 5.3\left(1+4\frac{\eta}{s}\right). 
\end{equation}
Higher viscosity results in a steeper distribution, and a measurement of the dilepton spectrum can provide an estimate of $\eta/s$. Intuitively, the inverse slope $T_{\rm inv}(M)=T_0+M/n$ of the distribution is the effective temperature  probed by dilepton production. 
High-mass dileptons are produced at earlier times and probe larger temperatures. 
Viscosity hinders early thermalization and results in lower effective temperatures. 
Note that $\eta/s$ denotes the effective viscosity at early times, which may differ from the effective viscosity measured through the analysis of anisotropic flow~\cite{JETSCAPE:2020mzn,Gardim:2020mmy}, which develops at later times. 

Based on excellent secondary vertexing, the large background from semileptonic charm and beauty hadron decays could be overcome with the next generation of heavy-ion experiments, LHCb U2 with dimuons and ALICE 3 with dielectrons at the LHC.
For precise performance assessments, detailed simulations of both detector set-ups for dilepton production in this mass range will be required and should enter into the detector design considerations. 

The work presented in this letter is based on state-of-the art knowledge of QCD kinetics and simplifies the space-time picture of heavy-ion collisions in order to keep the calculation compact and transparent. 
We have only studied the dependence of the dilepton yield on invariant mass $M$, but the transverse momentum spectrum of thermal dileptons can be calculated along the same lines. 
This type of calculation can be systematically improved from studying the dilepton production in QCD kinetics beyond leading order, to going beyond the transverse homogeneity assumption and treating the expansion dynamics in its full dimensionality.

\section*{Acknowledgments} 
We thank Ralf Rapp for sharing his calculation results at an early stage of our work, and Michael Strickland for useful comments. 
We thank Jessica Churchill for sharing the details of her calculations. 
We thank Giuliano Giacalone for evaluating the values of the transverse area $A$ in the \trento{} model. 
This work is supported in part by the Deutsche Forschungsgemeinschaft (DFG, German Research Foundation) through the CRC-TR 211 ’Strong-interaction matter under extreme conditions’– project number 315477589 – TRR 211 and in part in the framework of the GLUODYNAMICS project funded by the "P2IO LabEx (ANR-10-LABX-0038)"  in the framework "Investissements d'Avenir" (ANR-11-IDEX-0003-01) managed by the Agence Nationale de la Recherche (ANR), France.

\end{document}